\documentclass[journal]{IEEEtran}
\usepackage{amsmath,amssymb}
\usepackage{graphicx}
\usepackage{array}
\usepackage{booktabs}
\usepackage{multirow}
\usepackage{url}
\graphicspath{{pdf/}{photo/}}

\begin{document}

\title{LWMa: Distributed Deployment of an Angle-Delay-Domain Wireless Foundation Model for Multiple Tasks}
\author{Zhizhen Ren, Xudong Zhang, \textit{Graduate Student Member, IEEE}, Jingbo Tan, \textit{Member, IEEE}, and Jintao Wang, \textit{Fellow, IEEE}}
\maketitle

\begin{abstract}
The growing diversity of wireless services in sixth-generation (6G) wireless communication systems calls for reusable channel representations that support multiple tasks across different propagation environments. However, deploying such representations between user equipment and the base station in frequency-division duplexing systems faces uplink feedback constraints. This paper proposes a distributed framework based on an angle-delay-domain large wireless model (LWMa) for multiple tasks. LWMa exploits angle-delay structure and delay truncation to improve feature extraction and reduce computational cost. A scenario-specific network inspired by Perceiver IO (input-output) compresses embeddings for channel reconstruction tasks, while other tasks use a low-dimensional compact global embedding. Embedding reconstruction training followed by channel-quality-guided fine-tuning preserves information useful for downstream processing. Simulations demonstrate that the proposed method supports effective channel reconstruction and multiple tasks under limited feedback with reduced deployment complexity. Further evaluations show that the recovered embeddings generalize to additional tasks and support split inference.
\end{abstract}

\begin{IEEEkeywords}
Wireless foundation model, angle-delay domain, embedding compression, distributed inference.
\end{IEEEkeywords}

\section{Introduction}
\label{sec:introduction}

\subsection{Background and Motivation}
\label{subsec:motivation}

The integration of artificial intelligence (AI) into sixth-generation (6G) wireless systems is expected to support increasingly diverse wireless services. The International Mobile Telecommunications (IMT)-2030 framework identifies AI and communication as one of its usage scenarios~\cite{lwma_imt2030}. This evolution calls for learning models that can support different wireless tasks and adapt to diverse propagation environments. In particular, channel state information (CSI) contains information relevant to channel reconstruction, propagation-state identification, user localization, and beam selection. Learning reusable representations from CSI therefore provides a promising basis for supporting multiple functions within a common wireless processing framework.

Conventional deep learning approaches often develop a dedicated model for a particular task and a particular scenario~\cite{lwma_csi_overview,lwma_csinet}. Although such specialization can provide effective solutions, deploying separate models across tasks and scenarios requires repeated data collection, training, and model maintenance. Wireless foundation models offer an alternative by learning generalizable channel representations through pre-training on large-scale, multi-scenario, and feature-rich channel data. For example, the large wireless model (LWM) learns generalizable channel embeddings through self-supervised pre-training and reuses them for downstream tasks and diverse scenarios~\cite{lwma_lwm}. This paradigm allows task-specific heads to exploit a shared feature extractor, motivating the use of foundation models in wireless systems with diverse service requirements and limited channel data.

However, the availability of a reusable representation does not naturally meet the deployment requirements. In frequency-division duplexing (FDD) systems, downlink channel observations are obtained at the user equipment (UE), whereas functions such as multi-user beam selection are performed at the base station (BS)~\cite{lwma_csi_overview,lwma_distributed_precoding}. Information extracted at the UE must be transmitted to the BS over the feedback link, and the overhead becomes a major concern as the channel dimension increases~\cite{lwma_csi_overview,lwma_csinet}. When a foundation model extracts channel representations at the UE, its outputs must therefore be made available to downstream processing at the BS under limited feedback. This calls for a deployment framework that accounts for where channel information is acquired and how it is conveyed to the modules executing different tasks.

Designing such a framework requires matching feedback to the information needs of downstream tasks. With a shared foundation encoder, channel reconstruction requires detailed features across antennas and subcarriers, whereas line-of-sight (LoS)/non-line-of-sight (NLoS) identification, localization, and beam selection can use compact global representations. The resulting challenge is to select suitable feedback for each task and compress high-dimensional embedding sequences while retaining the information needed for downstream processing.

\subsection{Related Work}
\label{subsec:related_work}

Pre-trained models have been adapted to wireless tasks through both language-model adaptation and wireless-domain pre-training. LLM4CP introduces channel-specific preprocessing, embedding, and output modules to adapt a pre-trained large language model (LLM) to channel prediction while freezing most backbone parameters~\cite{lwma_llm4cp}. \cite{lwma_multitask_llm} uses task instructions and dedicated input and output modules with a shared LLM to support multi-user precoding, signal detection, and channel prediction, employing low-rank adaptation and quantization to reduce adaptation costs. These approaches adapt pre-trained language models to wireless tasks, whereas wireless-specific foundation models learn directly from channel data. This enables compact encoders tailored to channel structure, offering a route to lower model-storage and inference costs for UE deployment. Following the pre-training-and-adaptation paradigm, the method in~\cite{lwma_ssl_localization} pre-trains a channel encoder on unlabeled channel estimates through joint-embedding self-supervision, then adapts the learned representations to localization in new environments, improving performance when labeled data are limited. However, its evaluation focuses on localization, leaving broader multi-task reuse to be established. LWM instead adopts task-agnostic pre-training on large-scale channel data, using masked channel modeling to learn reusable embeddings that improve downstream classification over raw-channel inputs, particularly when labeled training data are limited~\cite{lwma_lwm}. CSI-CLIP exploits the consistency between CSI and channel impulse responses through contrastive learning, obtaining channel representations that support localization, beam management, and channel identification~\cite{lwma_csiclip}. The wireless world model (WWM) further incorporates environmental point clouds and user trajectories through multimodal latent prediction, improving performance over unimodal and task-specific baselines in its evaluated tasks, including unseen scenarios and real-world measurements~\cite{lwma_wwm}. However, in an FDD deployment where feature extraction and task execution reside at the UE and BS, respectively, directly transmitting detailed embedding sequences can impose substantial feedback overhead, making their efficient delivery a necessary part of model deployment.

Learning-based CSI feedback reduces uplink overhead through learned channel compression and reconstruction~\cite{lwma_csi_overview}. CsiNet learns a UE encoder and a BS reconstruction decoder~\cite{lwma_csinet}, while CRNet exploits channel features at multiple resolutions~\cite{lwma_crnet}. Beyond channel reconstruction, the joint design in~\cite{lwma_distributed_precoding} learns UE feedback and BS multi-user precoding together, optimizing the feedback for the final transmission objective. To improve generalization across propagation environments, UniversalNet standardizes CSI inputs and aligns their sparse-domain structure, allowing existing feedback models to accommodate distribution changes without updating their network weights~\cite{lwma_universalnet}. A complementary approach in~\cite{lwma_signal_compression} uses features extracted by a pre-trained wireless foundation model as additional inputs to a CSI compressor, improving channel reconstruction quality. Although these methods improve CSI reconstruction and generalization, they do not directly address the recovery of a foundation encoder's full embedding sequence for reuse by independently trained task heads.

Language-model compression provides complementary approaches to reducing representation size. AutoCompressors recursively summarize text segments into continuous vectors used as soft prompts~\cite{lwma_autocompressors}. Gist-token training restricts attention so that prompt information is concentrated into a small set of reusable token representations~\cite{lwma_gist}. The in-context autoencoder (ICAE) learns memory slots using text reconstruction and continuation objectives with a frozen language-model decoder~\cite{lwma_icae}. LLMLingua-2 instead uses distilled data and a bidirectional encoder to classify which text tokens should be retained~\cite{lwma_llmlingua2}. Although these studies preserve useful textual information with fewer tokens or memory slots, their compressed representations are designed for direct use by a language model, rather than for reconstructing the original embedding sequence. Moreover, their language-oriented objectives do not account for how embedding errors affect channel reconstruction. Adapting these ideas to wireless feedback therefore requires both an embedding recovery mechanism and a training objective that reflects channel reconstruction quality.

Our previous work introduced LWMa, an angle-delay (AD) domain variant of LWM, and an FDD framework for multi-user beam selection with low-dimensional embedding feedback~\cite{lwma_previous}. Extending that framework to channel feedback and interpolation requires recovery of detailed channel embeddings. This motivates task-oriented feedback with embedding recovery and channel-quality-guided compressor training.

\subsection{Contributions}
\label{subsec:contributions}

To address these challenges, this paper develops an LWMa-based framework for supporting multiple downstream tasks in FDD systems. A common pre-trained foundation encoder extracts channel representations at the UE. For channel feedback and interpolation, the UE compresses the encoder output using a scenario-specific compression network and transmits the resulting low-dimensional representation over the uplink. For LoS/NLoS identification, user localization, and single- and multi-user beam selection, the UE directly transmits the uncompressed classification (CLS) token embedding over the uplink. Task-specific heads are deployed at the BS. Within this framework, an angle-delay-domain architecture and a channel-quality-guided compression strategy are designed to support detailed channel reconstruction under limited feedback. The main contributions are summarized as follows:

\begin{itemize}
\item We develop a foundation model-based UE-BS deployment framework that supports channel feedback, channel interpolation, LoS/NLoS identification, user localization, and single- and multi-user beam selection. A scenario-specific compressor serves as a plug-in module shared by channel feedback and interpolation, while LoS/NLoS identification, user localization, and single- and multi-user beam selection reuse the foundation model through direct low-dimensional CLS feedback. By matching the feedback representation to task requirements, the framework makes a common foundation encoder available to diverse BS-side tasks under limited feedback.

\item We design LWMa, a wireless foundation architecture that represents channels in the angle-delay domain. This angle-delay domain adaptation, together with two-dimensional patch modeling and pixel masking, provides a basis for learning generalizable channel features through self-supervised pre-training. Delay-domain truncation further shortens the embedding sequence and reduces the computational cost of the foundation encoder.

\item We design an embedding compression and recovery network inspired by Perceiver IO (input-output)~\cite{lwma_perceiverio}. Learnable latent queries aggregate the embedding sequence into compact slots, which are projected into a low-dimensional feedback code. At the BS, a separate set of position-specific output queries recovers the token sequence through cross-attention to the decoded slots. After embedding reconstruction training, a separate fine-tuning stage incorporates channel reconstruction loss, with the reconstructed channel obtained by passing the recovered channel embeddings through the frozen pre-trained foundation decoder. This training strategy explicitly accounts for the channel quality associated with recovered embeddings.

\end{itemize}

The evaluation is organized around feedback efficiency, downstream task performance, and deployment complexity, with LWM-based and task-specific models as comparison points. It also examines the relationship between embedding recovery and channel reconstruction, providing a basis for assessing the proposed compression strategy across different feedback budgets.

\subsection{Organization and Notation}
\label{subsec:organization_notation}

The remainder of this paper is organized as follows. Section~\ref{sec:system_model} introduces the FDD system model and formulates the downstream tasks. Section~\ref{sec:framework} presents the LWMa-based deployment framework, including foundation pre-training, scenario-specific compressor adaptation, task-head training, and online inference. Section~\ref{sec:architecture_training} details the LWMa architecture and the embedding compression and training strategy. Section~\ref{sec:results} presents the simulation setup and performance evaluation. Finally, Section~\ref{sec:conclusion} concludes the paper.

\textit{Notation:} Bold uppercase and lowercase letters denote matrices and vectors, respectively, and ordinary letters denote scalars. The superscripts $(\cdot)^{\mathrm{T}}$ and $(\cdot)^{\mathrm{H}}$ represent transpose and Hermitian transpose, respectively. The symbols $|\cdot|$, $\|\cdot\|_2$, and $\|\cdot\|_{\mathrm{F}}$ denote the absolute value (or set cardinality), Euclidean norm, and Frobenius norm, respectively. The operator $\odot$ denotes element-wise multiplication, and $\oslash$ denotes element-wise division. The indicator function $\mathbb{I}\{\cdot\}$ equals one if its argument is true and zero otherwise. The operator $\mathbb{E}[\cdot]$ denotes expectation, and $\mathbf{I}$ denotes an identity matrix of the appropriate size. The sets of real and complex numbers are denoted by $\mathbb{R}$ and $\mathbb{C}$, respectively. $\mathcal{CN}(0,\sigma^2)$ denotes a zero-mean circularly symmetric complex Gaussian distribution with variance $\sigma^2$, and i.i.d. stands for independent and identically distributed.

\section{System Model and Downstream Tasks}
\label{sec:system_model}

This section introduces the FDD system model and defines the downstream tasks.

\subsection{FDD Massive MIMO-OFDM System}
\label{subsec:fdd_model}

In this paper, we consider a single-user FDD massive multiple-input multiple-output (MIMO) system employing orthogonal frequency-division multiplexing (OFDM) with $N_c$ subcarriers. The BS is equipped with $N_t$ transmit antennas, and the UE has a single receive antenna. The received signal on the $n$-th subcarrier is
\begin{equation}
y_n=\mathbf{h}_n^{\mathrm H}\mathbf{x}_n+w_n,
\qquad w_n\sim\mathcal{CN}(0,\sigma^2),
\label{eq:received_signal}
\end{equation}
where $\mathbf{x}_n\in\mathbb{C}^{N_t}$ is the transmitted signal, $w_n$ is additive white Gaussian noise with variance $\sigma^2$, and $\mathbf{h}_n\in\mathbb{C}^{N_t}$ is the downlink channel vector on subcarrier $n$. Stacking the channel vectors gives the space-frequency (SF) domain channel $\mathbf{H}=[\mathbf{h}_1,\ldots,\mathbf{h}_{N_c}]^{\mathrm H}
\in\mathbb{C}^{N_c\times N_t}$.
The AD domain channel is obtained as
\begin{equation}
\mathbf{H}^{\mathrm{AD}}=\mathbf{F}_d\mathbf{H}\mathbf{F}_a^{\mathrm H},
\label{eq:ad_channel}
\end{equation}
where $\mathbf{F}_d\in\mathbb{C}^{N_c\times N_c}$ and $\mathbf{F}_a\in\mathbb{C}^{N_t\times N_t}$ are Fourier transform matrices for the frequency-delay and angle-space transformations, respectively. Due to the limited delay spread, most channel energy is concentrated in the first $N_\tau$ rows. Retaining these rows yields the truncated AD domain channel $\widetilde{\mathbf{H}}^{\mathrm{AD}}\in\mathbb{C}^{N_\tau\times N_t}$, where $N_\tau<N_c$, reducing the input dimension while preserving the dominant channel components.

\subsection{Channel Feedback}
\label{subsec:channel_estimation}

Due to the non-reciprocity of uplink and downlink channels in FDD systems, the BS relies on UE feedback to acquire downlink channel information.
The UE obtains an initial estimate of the downlink SF domain channel and feeds back a representation from which the BS reconstructs the channel. The noisy estimated channel at the UE can be written as
\begin{equation}
{\mathbf{H}}^{\mathrm{fb}}=\mathbf{H}+\mathbf{N},
\qquad [\mathbf{N}]_{n,a}\mid\mathbf{H}\overset{\mathrm{i.i.d.}}{\sim}\mathcal{CN}(0,\sigma_{\mathrm{obs}}^2),
\label{eq:est_observation}
\end{equation}
where the observation-noise variance is $\sigma_{\mathrm{obs}}^2=\frac{\|\mathbf{H}\|_{\mathrm F}^2}{N_cN_t\rho_{\mathrm{obs}}}$.
$\rho_{\mathrm{obs}}$ is the linear observation signal-to-noise ratio (SNR). 

In the massive MIMO-OFDM system, the channel matrix is typically of high dimension. Directly feeding back all $2N_tN_c$ real-valued components of the complex channel incurs substantial feedback overhead. An autoencoder-based feedback scheme reduces this overhead by encoding the observed channel into a low-dimensional feedback vector and decoding it at the BS. These operations are written as
\begin{equation}
\mathbf{f}=\mathcal{E}_{\mathrm{fb}}({\mathbf{H}}^{\mathrm{fb}}),
\qquad \widehat{\mathbf{H}}^{\mathrm{fb}}=\mathcal{D}_{\mathrm{fb}}(\mathbf{f}),
\label{eq:feedback_interface}
\end{equation}
where $\mathcal{E}_{\mathrm{fb}}$ denotes the UE-side encoder, $\mathbf{f}\in\mathbb{R}^{d_f}$ is the complete feedback vector, including any auxiliary information required for reconstruction, with $d_f<2N_tN_c$, and $\mathcal{D}_{\mathrm{fb}}$ denotes the BS-side decoder. In this paper, we assume perfect feedback. Note that the channel compression and reconstruction can also be executed in the AD domain.

Reconstruction quality is measured by the normalized mean squared error (NMSE) of the SF domain channel,
\begin{equation}
\mathrm{NMSE}_{\mathrm{fb}}=\mathbb{E}\!\left[
\frac{\|\widehat{\mathbf{H}}^{\mathrm{fb}}-\mathbf{H}\|_{\mathrm F}^2}
{\|\mathbf{H}\|_{\mathrm F}^2}\right].
\label{eq:channel_nmse}
\end{equation}

\subsection{Channel Interpolation}
\label{subsec:channel_interpolation}

In MIMO-OFDM systems, acquiring full CSI across antennas and subcarriers can incur substantial measurement overhead, motivating channel recovery from sparse observations. Channel interpolation exploits spatial and frequency correlations to infer unobserved channel coefficients. 
The sparse observed channel is
\begin{equation}
{\mathbf{H}}^{\mathrm{int}}=\mathbf{M}\odot\mathbf{H},
\label{eq:interpolation_observation}
\end{equation}
 where unobserved entries are filled with zeros. We consider a cross-shaped sampling pattern with antenna stride $s_a$ and subcarrier stride $s_c$:
\begin{equation}
[\mathbf{M}]_{n,a}=\begin{cases}
1,& (a-1)\bmod s_a=0\ \text{or}\ (n-1)\bmod s_c=0,\\
0,& \text{otherwise}
\end{cases}.
\label{eq:cross_mask}
\end{equation}
Thus, the observed positions form the union of the selected antenna columns and subcarrier rows, including their intersections only once. The observation fraction is
\begin{equation}
p_{\mathrm{obs}}=\frac{N_a^{\mathrm{obs}}}{N_t}
+\frac{N_c^{\mathrm{obs}}}{N_c}
-\frac{N_a^{\mathrm{obs}}N_c^{\mathrm{obs}}}{N_tN_c},
\label{eq:observation_fraction}
\end{equation}
where $N_a^{\mathrm{obs}}=\lceil N_t/s_a\rceil$ and $N_c^{\mathrm{obs}}=\lceil N_c/s_c\rceil$. The target of channel interpolation is reconstructing the complete clean SF domain channel, which is evaluated using~\eqref{eq:channel_nmse} with $\widehat{\mathbf{H}}^{\mathrm{fb}}$ replaced by $\widehat{\mathbf{H}}^{\mathrm{int}}$.

\subsection{LoS/NLoS Identification}
\label{subsec:los_nlos}

LoS/NLoS identification determines whether an unobstructed direct path exists between the BS and UE, providing useful information for link adaptation and channel-based localization.
Let $\ell\in\{0,1\}$ denote the label, with $\ell=1$ for LoS and $\ell=0$ for NLoS. The BS predicts the LoS probability $\widehat{p}$ and assigns $\widehat{\ell}=\mathbb{I}\{\widehat{p}\geq 1/2\}$. The loss function is cross-entropy,
\begin{equation}
\mathcal{L}_{\mathrm{cls}}=-\mathbb{E}\!\left[
\ell\log\widehat{p}+(1-\ell)\log(1-\widehat{p})\right].
\label{eq:los_loss}
\end{equation}
Performance is evaluated by the identification accuracy $\mathbb{E}[\mathbb{I}\{\widehat{\ell}=\ell\}]$.

\subsection{User Localization}
\label{subsec:localization}

User localization estimates the two-dimensional UE position $\mathbf{r}=[x,y]^{\mathrm T}\in\mathbb{R}^2$ in the scenario coordinate system from location-dependent channel features, supporting location-aware communication and services.
To facilitate stable training and balance learning across spatial dimensions, we first normalize UE positions. Let $\boldsymbol{\mu}_{r}$ and $\boldsymbol{s}_{r}$ denote the mean and standard deviation computed from training positions. The localization head is trained by minimizing
\begin{equation}
\mathcal{L}_{\mathrm{loc}}=\mathbb{E}\!\left[
\|\widehat{\widetilde{\mathbf{r}}}-\widetilde{\mathbf{r}}\|_2^2\right],
\label{eq:localization_loss}
\end{equation}
where $\widetilde{\mathbf{r}}=(\mathbf{r}-\boldsymbol{\mu}_{r})\oslash\boldsymbol{s}_{r}$, and $\widehat{\widetilde{\mathbf{r}}}$ is the predicted normalized position.
The physical-coordinate prediction is $\widehat{\mathbf{r}}=\boldsymbol{\mu}_{r}+\boldsymbol{s}_{r}\odot\widehat{\widetilde{\mathbf{r}}}$, and the mean positioning error (MPE) is
\begin{equation}
\mathrm{MPE}=\mathbb{E}\!\left[
\|\widehat{\mathbf{r}}-\mathbf{r}\|_2\right].
\label{eq:position_error}
\end{equation}
The training statistics are retained at the BS for inference.

\subsection{Single- and Multi-User Beam Selection}
\label{subsec:beam_selection}

Beamforming is important for improving spectral efficiency in massive MIMO systems. Deep-learning-based beam selection aims to choose the best vector from a predefined codebook $\mathcal{F}=\{\mathbf{f}_1,\ldots,\mathbf{f}_Q\}$ to determine the analog beamforming vector while avoiding the training overhead of exhaustive beam sweeping, where $\mathbf{f}_q\in\mathbb{C}^{N_t}$ and $\|\mathbf{f}_q\|_2=1$. For a uniform linear array with antenna spacing $d$ and carrier wavelength $\lambda_c$, the codeword $\mathbf{f}_q$ is
\begin{equation}
\mathbf{f}_q=\frac{1}{\sqrt{N_t}}
[1,e^{\mathrm{j}2\pi d\sin\theta_q/\lambda_c},\ldots,
 e^{\mathrm{j}2\pi(N_t-1)d\sin\theta_q/\lambda_c}]^{\mathrm T},
\label{eq:beam_codebook}
\end{equation}
where $\theta_q=\theta_{\min}+\frac{q-1}{Q-1}(\theta_{\max}-\theta_{\min}),\quad q=1,\ldots,Q$.

\textit{Single-user beam selection:} For a single UE, the objective is to select the beam that maximizes its achievable spectral efficiency (SE). To enable stable training, we normalize the channel as
\begin{equation}
a=\sqrt{\frac{1}{N_c}\sum_{n=1}^{N_c}\|\mathbf{h}_{n}\|_2^2},
\qquad \overline{\mathbf{h}}_{n}=\mathbf{h}_{n}/a.
\label{eq:su_channel_scale}
\end{equation}
The mean spectral efficiency associated with beam $q$ is
\begin{equation}
R^{\mathrm{SU}}(q)=\frac{1}{N_c}\sum_{n=1}^{N_c}
\log_2\bigl(1+|\overline{\mathbf{h}}_{n}^{\mathrm H}\mathbf{f}_q|^2/\sigma^2\bigr),
\label{eq:su_beam_target}
\end{equation}
where $\sigma^2$ denotes the receiver noise variance. The goal of beam selection is to find the optimal beam index $q^\star=\arg\max_{q\in\{1,\ldots,Q\}}R^{\mathrm{SU}}(q)$. Top-1 and top-3 accuracy, the average achieved rate, and its ratio to the average oracle rate $\eta_{\mathrm{SU}}=
\frac{\mathbb{E}[R^{\mathrm{SU}}(\widehat q)]}
{\mathbb{E}[R^{\mathrm{SU}}(q^\star)]}$ are reported.

\textit{Multi-user beam selection:} When multiple UEs share the same time-frequency resources, independently selected beams may cause inter-user interference, motivating joint beam selection to improve the sum spectral efficiency. The single-user setting can be extended to a scheduled group $\mathcal{U}=\{1,\ldots,K\}$ of $K$ single-antenna UEs served simultaneously by the BS. We consider a hybrid beamforming architecture with one radio-frequency (RF) chain per UE. The received signal at UE $k\in\mathcal{U}$ on subcarrier $n$ is
\begin{equation}
y_{k,n}=\mathbf{h}_{k,n}^{\mathrm H}\mathbf{F}_{\mathrm{RF}}\mathbf{F}_{\mathrm{BB},n}\mathbf{s}_n+w_{k,n},
\qquad w_{k,n}\sim\mathcal{CN}(0,\sigma^2),
\label{eq:mu_received_signal}
\end{equation}
where $\mathbf{h}_{k,n}\in\mathbb{C}^{N_t}$ is the SF domain channel vector of UE $k$ on subcarrier $n$, $\mathbf{s}_n\in\mathbb{C}^{K}$ is the transmitted symbol vector with $\mathbb{E}[\mathbf{s}_n\mathbf{s}_n^{\mathrm H}]=\mathbf{I}_K$, and $\mathbf{F}_{\mathrm{RF}}\in\mathbb{C}^{N_t\times K}$ and $\mathbf{F}_{\mathrm{BB},n}=[\mathbf{f}_{\mathrm{BB},1,n},\ldots,\mathbf{f}_{\mathrm{BB},K,n}]\in\mathbb{C}^{K\times K}$ denote the analog beamformer and digital precoder, respectively. For a beam-index vector $\mathbf{q}=[q_1,\ldots,q_K]^{\mathrm T}$, the analog beamformer is $\mathbf{F}_{\mathrm{RF}}(\mathbf{q})=[\mathbf{f}_{q_1},\ldots,\mathbf{f}_{q_K}]$. Given the selected analog beams, digital precoding like zero-forcing can be designed from the effective channel. The precoders satisfy $\|\mathbf{F}_{\mathrm{RF}}\mathbf{F}_{\mathrm{BB},n}\|_{\mathrm F}^2=K$, corresponding to unit average power per stream.

Similarly, a common group normalization preserves the relative channel strengths:
\begin{equation}
\check{\mathbf{h}}_{k,n}=\alpha_{\mathcal{U}}\mathbf{h}_{k,n},\qquad
\alpha_{\mathcal{U}}=\sqrt{\frac{K}{\sum_{v\in\mathcal{U}}\sum_{n=1}^{N_c}
\|\mathbf{h}_{v,n}\|_2^2}}.
\label{eq:mu_channel_scale}
\end{equation}
The signal-to-interference-plus-noise ratio is
\begin{equation}
\gamma_{k,n}(\mathbf{q})=
\frac{|\check{\mathbf{h}}_{k,n}^{\mathrm H}\mathbf{F}_{\mathrm{RF}}(\mathbf{q})\mathbf{f}_{\mathrm{BB},k,n}|^2}
{\sum_{v\in\mathcal{U},v\ne k}|\check{\mathbf{h}}_{k,n}^{\mathrm H}\mathbf{F}_{\mathrm{RF}}(\mathbf{q})\mathbf{f}_{\mathrm{BB},v,n}|^2+\sigma^2}.
\label{eq:mu_sinr}
\end{equation}
The sum spectral efficiency of the multi-user system is
\begin{equation}
R_{\mathcal{U}}^{\mathrm{MU}}(\mathbf{q})
=\frac{1}{N_c}\sum_{k\in\mathcal{U}}\sum_{n=1}^{N_c}\log_2(1+\gamma_{k,n}(\mathbf{q})).
\label{eq:mu_beam_target}
\end{equation}
The goal is to choose the optimal group of beam indices $\mathbf{q}_{\mathcal{U}}^\star
=\arg\max_{\mathbf{q}\in\{1,\ldots,Q\}^{K}}R_{\mathcal{U}}^{\mathrm{MU}}(\mathbf{q})$.
The normalized spectral efficiency is also evaluated as $
\eta_{\mathrm{MU}}=
\frac{\mathbb{E}_{\mathcal{U}}[R_{\mathcal{U}}^{\mathrm{MU}}(\widehat{\mathbf{q}}_{\mathcal{U}})]}
{\mathbb{E}_{\mathcal{U}}[R_{\mathcal{U}}^{\mathrm{MU}}(\mathbf{q}_{\mathcal{U}}^\star)]}$.


\begin{figure*}[!t]
\centering
\includegraphics[width=\textwidth]{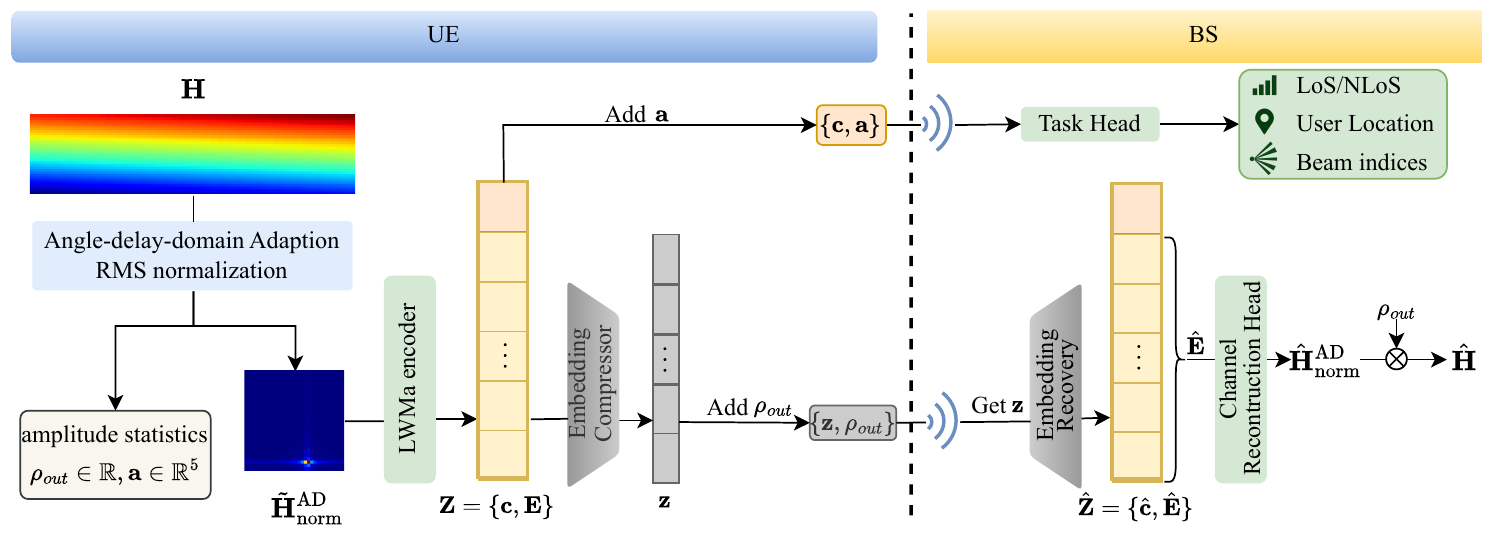}
\caption{The proposed framework of deployment.}
\label{fig:framework}
\end{figure*}

\section{LWMa-Based Distributed Framework for Multiple Tasks}
\label{sec:framework}

To support the tasks in Section~\ref{sec:system_model}, task-specific models can be trained separately for each deployment scenario. However, this approach requires repeated model training with varying tasks and environments, and suffers from poor generalization in unseen scenarios. Therefore, we develop an LWMa-based distributed framework that shares a common channel encoder across tasks and scenarios. This framework combines generalizable channel representations with task-oriented feedback, enabling the UE to extract features from its locally observed downlink channel and the BS to perform downstream inference in FDD systems. Its offline preparation comprises foundation-model pre-training, task-oriented embedding compression and feedback, and task-head training, followed by UE-BS deployment.

\subsection{Shared LWMa Foundation Model Pre-training}
\label{subsec:pre-training_workflow}

First, the shared LWMa foundation model is pre-trained on a large-scale dataset including multiple scenarios to enable generalization.
Let $\mathcal{F}_{\mathrm{enc}}(\cdot;\boldsymbol{\Theta}_{\mathrm{enc}})$ and $\mathcal{F}_{\mathrm{dec}}(\cdot;\boldsymbol{\Theta}_{\mathrm{dec}})$ denote the LWMa encoder and its pre-training reconstruction decoder, respectively. 
By reconstructing randomly masked channel components in input channels in a self-supervised approach, the encoder learns channel features that can be used for different downstream tasks. The network architecture and pre-training objective are presented in Section~\ref{subsec:lwma_architecture}.

For a channel sample, let $\mathbf{X}$ denote the input patch sequence obtained from the available channel observation through AD domain preprocessing and patch embedding. The encoder produces
\begin{equation}
\mathbf{Z}=\{\mathbf{c},\mathbf{E}\}
=\mathcal{F}_{\mathrm{enc}}(\mathbf{X};\boldsymbol{\Theta}_{\mathrm{enc}}),
\label{eq:framework_representation}
\end{equation}
where $\mathbf{Z}=[\mathbf{c},\mathbf{e}_{1},\ldots,\mathbf{e}_{P}]^{\mathrm T}\in\mathbb{R}^{(P+1)\times D}$ denotes the complete sequence; $\mathbf{c}\in\mathbb{R}^{D}$ is the CLS embedding and $\mathbf{E}=[\mathbf{e}_{1},\ldots,\mathbf{e}_{P}]^{\mathrm T}\in\mathbb{R}^{P\times D}$ contains the $P$ channel-patch embeddings of dimension $D$. The CLS embedding summarizes global channel information, whereas the channel embeddings retain detailed channel features.
The decoder enables self-supervised pre-training and is retained as a fixed channel-reconstruction network for embedding compressor refinement, as described in Section~\ref{subsec:compression_training}. It is discarded during downstream inference.

\subsection{Task-Oriented Embedding Feedback}
\label{subsec:scenario_compression}

The generalization capability of the pre-trained LWMa enables adaptation to downstream tasks in scenarios not encountered during pre-training. 
Nevertheless, a task-oriented embedding feedback scheme is designed to satisfy the feedback constraints of FDD systems.
In the proposed framework, channel feedback and interpolation use the detailed channel embeddings $\mathbf{E}$ as the task-head input, whereas LoS/NLoS identification, localization, and beam selection use the global CLS embedding $\mathbf{c}$. While the low-dimensional CLS embedding can be fed back directly, the high dimension of $\mathbf{E}$ makes its transmission costly. Therefore, we introduce compression and recovery networks to reduce the feedback overhead of $\mathbf{E}$.

Let $\mathcal{F}_{\mathrm{com}}(\cdot;\boldsymbol{\Theta}_{\mathrm{com}})$ and $\mathcal{F}_{\mathrm{rec}}(\cdot;\boldsymbol{\Theta}_{\mathrm{rec}})$ denote the compression and recovery networks associated with a target scenario. The compressor takes the complete embedding sequence, including the CLS token, as input. The feedback process is described as
\begin{equation}
\mathbf{z}=\mathcal{F}_{\mathrm{com}}(\mathbf{Z};\boldsymbol{\Theta}_{\mathrm{com}}),
\widehat{\mathbf{Z}}=\mathcal{F}_{\mathrm{rec}}(\mathbf{z};\boldsymbol{\Theta}_{\mathrm{rec}}),
\label{eq:framework_compression}
\end{equation}
where $\mathbf{z}\in\mathbb{R}^{d_z}$ with $d_z<(P+1)D$ is the compressed code, and $\widehat{\mathbf{c}}\in\mathbb{R}^{D}$ and $\widehat{\mathbf{E}}\in\mathbb{R}^{P\times D}$ are the recovered CLS and channel embeddings in $\widehat{\mathbf{Z}}$, respectively. Although the CLS token is not used by the channel estimation and interpolation heads, it supplies global context and is an integral part of the network input. 
The compression and recovery network serves as a scenario-specific plug-in module shared by channel feedback and interpolation. Networks with different $d_z$ can be selected to match the feedback budget.


\subsection{Task-Specific Head Adaptation}
\label{subsec:head_training}

To support multiple downstream tasks with the unified LWMa encoder, task-specific heads are trained separately and deployed at the BS.
During head training, the encoder generates representations from the channel observations, and only the task head parameters are optimized. 

The heads for channel feedback and interpolation are trained on the original, uncompressed channel embeddings $\mathbf{E}$ without considering the embedding compression and recovery network.
The heads for LoS/NLoS identification, localization, and beam selection are trained using the original CLS embeddings $\mathbf{c}$ together with amplitude statistics and their task labels, such as propagation states, UE positions, or reference beam indices.

\subsection{UE-BS Deployment and Online Inference}
\label{subsec:deployment}

During online inference, the complete LWMa encoder is deployed at the UE. The UE also stores the compression network associated with the target scenario, while the BS stores the corresponding recovery network and task-specific heads. The plug-in design of the compression and recovery network enables flexible deployment selection according to channel conditions and available feedback resources. After acquiring the channel observation, the UE performs the required preprocessing and extracts $\mathbf{Z}$ using the encoder.

For channel feedback or interpolation, the UE compresses $\mathbf{Z}$ into $\mathbf{z}\in\mathbb{R}^{d_z}$ and feeds back $\mathbf{f}=[\mathbf{z}^{\mathrm T},\rho_{\mathrm{out}}]^{\mathrm T}\in\mathbb{R}^{d_z+1}$, where $\rho_{\mathrm{out}}$ is the scale required to restore the channel amplitude. The BS recovers $\widehat{\mathbf{Z}}$ from $\mathbf{z}$ and supplies $\widehat{\mathbf{E}}$ to the selected head:
\begin{equation}
\widehat{\mathbf{H}}_{\mathrm{norm}}^{\mathrm{AD,fb}}
=\mathcal{F}_{\mathrm{fb}}(\widehat{\mathbf{E}};\boldsymbol{\Theta}_{\mathrm{fb}}),
\widehat{\mathbf{H}}_{\mathrm{norm}}^{\mathrm{AD,int}}
=\mathcal{F}_{\mathrm{int}}(\widehat{\mathbf{E}};\boldsymbol{\Theta}_{\mathrm{int}}),
\label{eq:framework_channel_inference}
\end{equation}
where $\mathcal{F}_{\mathrm{fb}}(\cdot;\boldsymbol{\Theta}_{\mathrm{fb}})$ and $\mathcal{F}_{\mathrm{int}}(\cdot;\boldsymbol{\Theta}_{\mathrm{int}})$ denote the heads for channel feedback and interpolation, respectively.
The normalized AD reconstruction is multiplied by $\rho_{\mathrm{out}}$, zero-padded in delay, and transformed back to the SF domain. The task-dependent scale is computed from the UE observation as specified in Section~\ref{subsec:task_head_architecture}.

For LoS/NLoS identification, localization, or beam selection, the UE feeds back $\mathbf{f}=[\mathbf{c}^{\mathrm T},\mathbf{a}^{\mathrm T}]^{\mathrm T}\in\mathbb{R}^{D+5}$, where $\mathbf{a}\in\mathbb{R}^{5}$ contains amplitude statistics of the unnormalized retained AD observation. The BS uses both quantities in the corresponding task head. These statistics are defined in Section~\ref{subsec:task_head_architecture}. For multi-user beam selection, the BS collects the representations of the scheduled UEs before predicting their beam indices jointly. The deployment framework is illustrated in Fig.~\ref{fig:framework}.

\section{Network Architecture and Training Strategy}
\label{sec:architecture_training}

To extract physically meaningful channel features for diverse downstream tasks, we design LWMa around the angular and delay structure of wireless propagation. To make the high-dimensional representations generated by the LWMa encoder at the UE available at the BS under limited feedback, we develop an embedding compressor that aggregates the full token sequence. A two-stage training strategy combines embedding reconstruction with channel-quality-guided refinement, encouraging the compressed representation to preserve information needed for channel recovery. This section details these network designs, their training objectives, and the task heads.

\subsection{Angle-Delay-Domain LWMa Architecture and Pretraining}
\label{subsec:lwma_architecture}

LWMa is designed to support generalizable channel feature extraction with a reduced input dimension for UE-side inference. Through AD domain adaptation, root-mean-square (RMS) normalization, pixel-level masking, and positional encoding, LWMa encourages efficient AD domain channel feature learning. The detailed design is described below.

\begin{figure*}[!t]
\centering
\includegraphics[width=\textwidth]{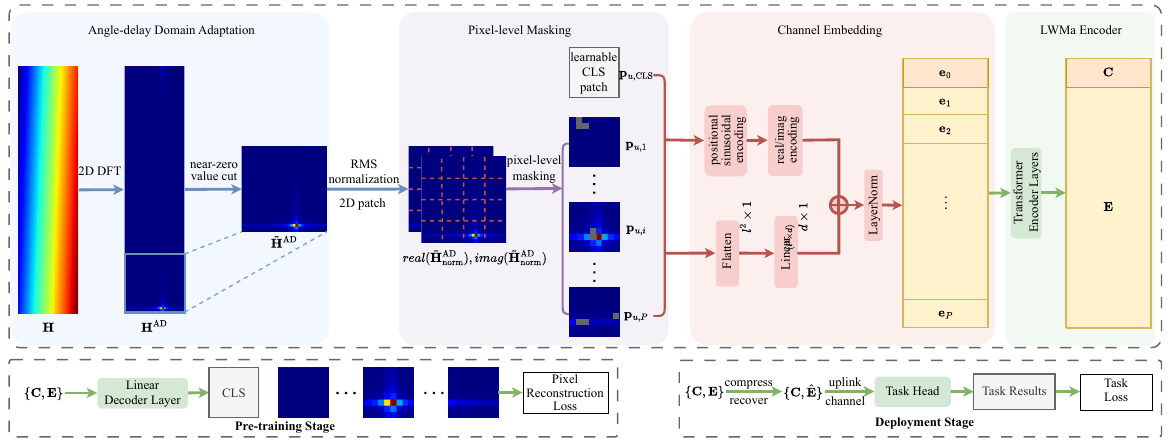}
\caption{The architecture of the proposed LWMa network.}
\label{fig:LWMa}
\end{figure*}

\textbf{Angle-delay domain adaptation:}
To make the physical relationships underlying the channel more explicit for feature learning, we first transform the SF domain channel into the AD domain using~\eqref{eq:ad_channel}, so that channel components are organized by propagation direction and delay. We then retain the first $N_\tau$ delay bins to obtain $\widetilde{\mathbf{H}}^{\mathrm{AD}}$, preserving the dominant components while reducing the channel dimension and computational complexity. Finally, we partition the real and imaginary arrays into non-overlapping $l\times l$ patches, allowing each token to describe a local angular-delay region. There are $P=2N_tN_\tau/l^2$ channel patches, each containing $L=l^2$ real-valued elements.

\textbf{RMS normalization:} Channel amplitudes can vary substantially across scenarios, causing large-amplitude samples to dominate reconstruction-based learning. To place samples on a common amplitude scale while preserving their internal channel structure, we normalize each clean pre-training sample according to
\begin{equation}
\rho_{\mathrm{in}}=2\sqrt{\frac{\|\widetilde{\mathbf{H}}^{\mathrm{AD}}\|_{\mathrm F}^2}{N_tN_\tau}},
\widetilde{\mathbf{H}}_{\mathrm{norm}}^{\mathrm{AD}}
=\frac{\widetilde{\mathbf{H}}^{\mathrm{AD}}}{\rho_{\mathrm{in}}}.
\label{eq:architecture_rms}
\end{equation}
This normalization preserves the complex phase and relative amplitudes within each channel and sets its complex RMS magnitude to $1/2$ to enable effective training.

\textbf{Pixel-level masking:} To provide a compact summary of the channel for downstream tasks, a learnable CLS input patch is added to the beginning of the channel-patch sequence to learn a global representation.
To enable self-supervised pre-training, masking channel components at the patch level and reconstructing the masked values are common approaches.
In the sparse AD domain, however, high-energy coefficients may occupy only a few patches, so masking entire patches can remove a large fraction of the informative channel components. We instead apply pixel-level masking to distribute reconstruction targets within patches. Specifically, $15\%$ of the complex channel positions are selected, together with their corresponding real and imaginary components. Each selected real-valued component is replaced by the mask value with probability $0.8$, by a random value with probability $0.1$, and left unchanged otherwise.

\textbf{Positional encoding:} To distinguish where a patch lies in the angular-delay grid and whether it represents the real or imaginary component, we introduce separate spatial and type encodings excluding the CLS patch. Both are normalized and assigned small weights, so that positional information is provided with a controlled contribution to the channel embedding. Let $(a_i,d_i)$ denote the angular and delay patch coordinates of token $i$, and let $\chi_i\in\{\mathrm{Re},\mathrm{Im}\}$ identify its component type. The spatial encoding is formed by concatenating two sinusoidal encodings and normalizing their joint norm:
\begin{equation}
\begin{aligned}
\mathbf{p}_i&=\frac{[\boldsymbol{\phi}(a_i)^{\mathrm T},
\boldsymbol{\phi}(d_i)^{\mathrm T}]^{\mathrm T}}
{\|[\boldsymbol{\phi}(a_i)^{\mathrm T},
\boldsymbol{\phi}(d_i)^{\mathrm T}]^{\mathrm T}\|_2},\\
[\boldsymbol{\phi}(k)]_{2j}&=\sin(k/10000^{4j/D}),\\
[\boldsymbol{\phi}(k)]_{2j+1}&=\cos(k/10000^{4j/D}),
\end{aligned}
\label{eq:adpl_position}
\end{equation}
where $\boldsymbol{\phi}(k)\in\mathbb{R}^{D/2}$ and $j=0,\ldots,D/4-1$. Real and imaginary patches at the same spatial position share the spatial encoding $\mathbf{p}_i$. A separate trainable type vector $\mathbf{t}_{\chi_i}\in\mathbb{R}^D$ distinguishes them. For the vectorized input patch $\mathbf{x}_{i}\in\mathbb{R}^{L}$, the initial embedding is
\begin{equation}
\mathbf{e}_{i}^{(0)}=\mathrm{LN}\!\left(
\mathbf{W}_{\mathrm{emb}}\mathbf{x}_{i}+\mathbf{b}_{\mathrm{emb}}
+\beta_{\mathrm p}\mathbf{p}_i
+\gamma_{\mathrm t}\frac{\mathbf{t}_{\chi_i}}{\|\mathbf{t}_{\chi_i}\|_2}
\right),
\label{eq:adpl_embedding}
\end{equation}
where $\mathrm{LN}$ denotes layer normalization, $\mathbf{W}_{\mathrm{emb}}\in\mathbb{R}^{D\times L}$ is a learned linear projection, and $\beta_{\mathrm p}$ and $\gamma_{\mathrm t}$ are the spatial and type weights, respectively. 
This positional encoding retains explicit spatial information while limiting its perturbation of the projected channel features.


\textbf{Self-supervised training:} To learn dependencies beyond individual patches, the resulting sequence passes through the Transformer encoder to produce $\{\mathbf{c},\mathbf{E}\}$ in~\eqref{eq:framework_representation}. Each encoder layer applies residual multi-head self-attention, layer normalization, and a feedforward network. A lightweight decoder reconstructs the channel components from the refined tokens in a self-supervised way. The decoding operation on each channel token is
\begin{equation}
\widehat{\mathbf{x}}_{i}
=\mathbf{W}_{\mathrm{out}}\mathrm{LN}\!\left(
\mathrm{ReLU}(\mathbf{W}_{\mathrm{pre}}\mathbf{e}_{i}+\mathbf{b}_{\mathrm{pre}})
\right)+\mathbf{b}_{\mathrm{out}}.
\label{eq:foundation_patch_decoder}
\end{equation}
Here, $\mathrm{ReLU}$ denotes the rectified linear unit activation. The decoded patches are assembled into the real and imaginary channel arrays. Let $\mathcal{M}_{\mathcal{B}}$ collect all selected real-valued target components in a training batch $\mathcal{B}$, and let $v_{\mathcal{B}}$ be their sample variance. The pre-training objective is
\begin{equation}
\mathcal{L}_{\mathrm{pre}}=
\frac{1}{v_{\mathcal{B}}|\mathcal{M}_{\mathcal{B}}|}
\sum_{m\in\mathcal{M}_{\mathcal{B}}}(\widehat{x}_m-x_m)^2.
\label{eq:pre-training_mcm}
\end{equation}
Restricting the reconstruction objective to the selected components encourages the encoder to infer missing channel information from its context.


\subsection{Embedding Compression and Recovery Architecture}
\label{subsec:compressor_architecture}

To meet the restriction of uplink feedback mentioned in Section~\ref{subsec:scenario_compression}, the embedding compression and recovery architecture must retain compact information from a long embedding sequence $\mathbf{Z}$ while recovering the token structure required by the downstream heads. Therefore, we design a Perceiver IO-inspired embedding compression and recovery network~\cite{lwma_perceiverio}. 
A small set of learnable latent slots aggregates information from the full embedding sequence. At the BS, a learned query for each original token position recovers the corresponding embedding from the decoded latent slots. Different compression budgets can therefore be accommodated while preserving the dimensions of the recovered embeddings supplied to the task head.

\textbf{Embedding compression at the UE:} 
%

To aggregate its information within a limited feedback budget, the UE generates $S$ learnable latent queries:
\begin{equation}
\mathbf{A}=\mathcal{F}_{\mathrm{com}}^{\mathrm{att}}
(\mathbf{Q}_{\mathrm s},\overline{\mathbf{Z}};
\boldsymbol{\Theta}_{\mathrm{com}}^{\mathrm{att}}),
\label{eq:slot_aggregation}
\end{equation}
where $\mathbf{Q}_{\mathrm s}\in\mathbb{R}^{S\times D}$ contains the learnable slots with slot-position vectors added, and $\overline{\mathbf{Z}}$ is the complete input sequence augmented with learned input-position embeddings. $\mathcal{F}_{\mathrm{com}}^{\mathrm{att}}(\cdot, \cdot; \boldsymbol{\Theta}_{\mathrm{com}}^{\mathrm{att}})$ is an attention module that uses the slots $\mathbf{Q}_s$ as queries and $\overline{\mathbf{Z}}$ as both keys and values in cross-attention blocks, each followed by self-attention among the slots. The resulting slots $\mathbf{A}\in\mathbb{R}^{S\times D}$ are projected into the code $\mathbf{z}\in\mathbb{R}^{d_z}$ with $d_z=Sr$:
\begin{equation}
\mathbf{z}=
\mathcal{F}_{\mathrm{com}}^{\mathrm{down}}
(\mathbf{A};\boldsymbol{\Theta}_{\mathrm{com}}^{\mathrm{down}}),
\label{eq:slot_dimension_reduction}
\end{equation}
where $\mathcal{F}_{\mathrm{com}}^{\mathrm{down}}
(\cdot;\boldsymbol{\Theta}_{\mathrm{com}}^{\mathrm{down}})$ applies a linear projection from $D$ to $r$ dimensions to each slot. $\mathcal{F}_{\mathrm{com}}^{\mathrm{att}}(\cdot, \cdot; \boldsymbol{\Theta}_{\mathrm{com}}^{\mathrm{att}})$ and $\mathcal{F}_{\mathrm{com}}^{\mathrm{down}}
(\cdot;\boldsymbol{\Theta}_{\mathrm{com}}^{\mathrm{down}})$ are both part of $\mathcal{F}_{\mathrm{com}}(\cdot;\boldsymbol{\Theta}_{\mathrm{com}})$ in~\eqref{eq:framework_compression}.

\textbf{Embedding recovery at the BS:} The BS reshapes the compressed code into $S$ slots and decodes them to the original feature width $D$:
\begin{equation}
\widetilde{\mathbf{A}}=\mathcal{F}_{\mathrm{rec}}^{\mathrm{up}}
\left(\mathbf{z};
\boldsymbol{\Theta}_{\mathrm{rec}}^{\mathrm{up}}\right),
\label{eq:slot_dimension_expansion}
\end{equation}
where $\mathcal{F}_{\mathrm{rec}}^{\mathrm{up}}$ applies a linear projection from $r$ to $D$ dimensions to each slot. Learnable output queries then recover the complete sequence:
\begin{equation}
\widehat{\mathbf{Z}}=\mathcal{F}_{\mathrm{rec}}^{\mathrm{att}}
(\mathbf{Q}_{\mathrm o},\overline{{\mathbf{A}}};
\boldsymbol{\Theta}_{\mathrm{rec}}^{\mathrm{att}}),
\label{eq:token_sequence_recovery}
\end{equation}
where $\mathbf{Q}_{\mathrm o}\in\mathbb{R}^{(P+1)\times D}$ contains the output-position queries. $\overline{{\mathbf{A}}}$ denotes the latent slots augmented with the slot-position vectors. $\mathcal{F}_{\mathrm{rec}}^{\mathrm{att}}
(\cdot,\cdot;
\boldsymbol{\Theta}_{\mathrm{rec}}^{\mathrm{att}})$ is a similar attention module to $\mathcal{F}_{\mathrm{com}}^{\mathrm{att}}(\cdot, \cdot; \boldsymbol{\Theta}_{\mathrm{com}}^{\mathrm{att}})$ using $\overline{{\mathbf{A}}}$ as keys and values for the output queries $\mathbf{Q}_{\mathrm o}$. 
The up-projection network $\mathcal{F}_{\mathrm{rec}}^{\mathrm{up}}$ and the attention module $\mathcal{F}_{\mathrm{rec}}^{\mathrm{att}}$ together implement $\mathcal{F}_{\mathrm{rec}}$ in~\eqref{eq:framework_compression}. The output $\widehat{\mathbf{Z}}$ contains $\widehat{\mathbf{c}}$ and $\widehat{\mathbf{E}}$. Although only $\widehat{\mathbf{E}}$ is supplied to the channel feedback and interpolation heads, the CLS embedding is also compressed and recovered, as it captures global channel information that facilitates the recovery of $\mathbf{E}$.

The compressed code $\mathbf{z}$ contains $d_z=Sr$ real-valued entries. The compression ratios relative to the full embedding and the original complex SF domain channel are
$\eta_{\mathrm{emb}}=\frac{Sr}{(P+1)D}$ and $\eta_{\mathrm{CSI}}=\frac{Sr}{2N_tN_c}$, respectively. These ratios exclude the auxiliary amplitude information.

\subsection{Embedding Reconstruction and Channel-Quality-Guided Fine-Tuning}
\label{subsec:compression_training}

Embedding reconstruction training encourages the recovered embedding to approximate the original encoder outputs. However, accurate embedding reconstruction does not necessarily imply accurate channel reconstruction. Therefore, we propose an embedding reconstruction and channel-quality-guided fine-tuning method. First, the compressor is trained to preserve the channel embedding, and then it is refined with a channel-quality objective. 
Let $\mathbf{z}_{i}$ and $\widehat{\mathbf{z}}_{i}$ denote token $i$ of $\mathbf{Z}$ and $\widehat{\mathbf{Z}}$, with $i=0$ corresponding to CLS. The embedding reconstruction loss of the first stage combines element-wise squared error and token-wise cosine dissimilarity:
\begin{equation}
\mathcal{L}_{\mathrm{emb}}
=\frac{1}{P+1}\mathbb{E}\!\big[\sum_{i=0}^{P}(
\frac{\|\widehat{\mathbf{z}}_{i}-\mathbf{z}_{i}\|_2^2}{D}
+\omega_{\mathrm{cos}}(1-
\frac{\widehat{\mathbf{z}}_{i}^{\mathrm T}\mathbf{z}_{i}}
{\|\widehat{\mathbf{z}}_{i}\|_2\|\mathbf{z}_{i}\|_2}))\big],
\label{eq:embedding_training_loss}
\end{equation}
where $\omega_{\mathrm{cos}}$ is the weight of cosine dissimilarity. The squared-error term constrains embedding values, while the cosine term encourages preservation of their directions.

Embedding similarity alone does not directly constrain the channel reconstruction quality from those embeddings. Therefore, in the fine-tuning stage we pass $\widehat{\mathbf{E}}$ through the fixed pre-trained decoder $\mathcal{F}_{\mathrm{dec}}$ and obtain the reconstructed truncated AD domain channel $\widehat{{\mathbf{H}}}^{\mathrm{AD}}$. The decoder computation retains gradients with respect to its input, allowing the channel-quality loss to update both compression and recovery networks. The channel reconstruction loss is
\begin{equation}
\mathcal{L}_{\mathrm{ch}}=\mathbb{E}\!\left[
\frac{\|\widehat{{\mathbf{H}}}^{\mathrm{AD}}-\widetilde{\mathbf{H}}^{\mathrm{AD}}
\|_{\mathrm F}^{2}}
{\|\widetilde{\mathbf{H}}^{\mathrm{AD}}\|_{\mathrm F}^{2}}
\right].
\label{eq:compressor_channel_loss}
\end{equation}
The total refinement loss in the fine-tuning stage is
\begin{equation}
\mathcal{L}_{\mathrm{ft}}=\mu\mathcal{L}_{\mathrm{emb}}+
\lambda\mathcal{L}_{\mathrm{ch}},
\label{eq:compressor_refinement_loss}
\end{equation}
where $\mu$ and $\lambda$ balance preservation of the embedding with the accuracy of the decoded channel.

\begin{figure}[!t]
\centering
\includegraphics[width=\linewidth]{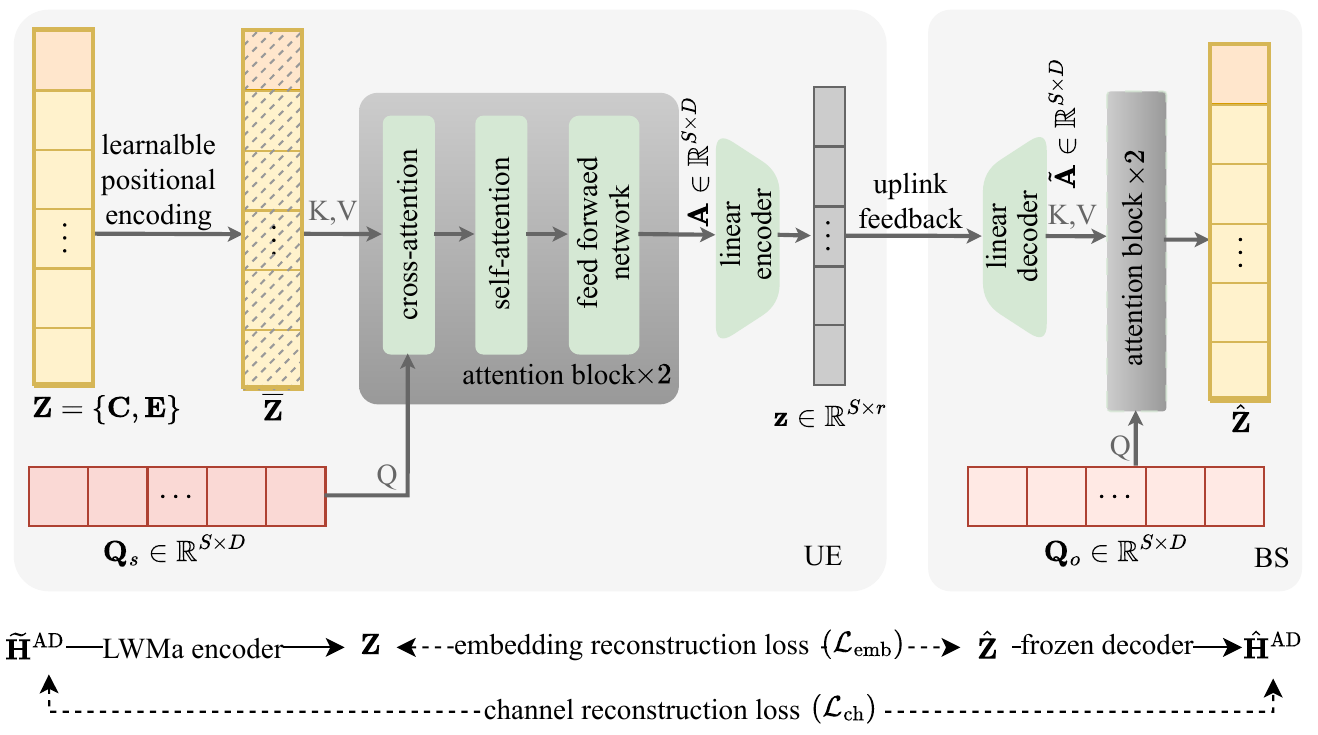}
\caption{The architecture of the proposed Perceiver IO-inspired embedding compression and recovery network.}
\label{fig:perceiver}
\end{figure}


\subsection{Downstream Task Heads and Optimization}
\label{subsec:task_head_architecture}

To adapt the universal embeddings generated by the LWMa encoder to different downstream tasks, we use task-specific heads trained according to Section~\ref{subsec:head_training}. The task-specific details of network architecture and optimization are given below.

\textbf{Channel feedback:} To keep input and target channel amplitudes consistent without requiring the clean channel scale, we normalize both using statistics of the noisy observation. Let ${\widetilde{\mathbf{H}}}^{\mathrm{AD,fb}}$ be the retained AD domain channel obtained from the noisy SF observation in~\eqref{eq:est_observation}. Let $\rho_{\mathrm{in}}$ denote the normalization scale of ${\widetilde{\mathbf{H}}}^{\mathrm{AD,fb}}$ computed using~\eqref{eq:architecture_rms}. Both the noisy input and clean AD target are divided by $\rho_{\mathrm{in}}$, so the output scale is $\rho_{\mathrm{out}}=\rho_{\mathrm{in}}$.

The normalized channel is passed through the LWMa encoder and the embedding compression and recovery network, forming the reconstructed embedding. We design a U-Net-based decoder to obtain the clean channel from the embedding. Firstly, a decoder using the structure in~\eqref{eq:foundation_patch_decoder} is applied to get the coarse channel. A residual U-Net then exploits neighboring angular-delay structure to refine this coarse reconstruction. The skip connection in the U-Net concatenates the encoder features with the upsampled features, and the predicted residual is added to the coarse grid. 

\textbf{Channel interpolation:} The same head structure is trained independently for channel interpolation to infer missing coefficients from the masked observations in~\eqref{eq:interpolation_observation}. Since channel masking reduces the observed energy and can redistribute it across delay bins, using the input normalization scale directly for the clean target can introduce an amplitude mismatch. Therefore, we use distinct input and output normalization scales. Let $\rho_{\mathrm{in}}$ be the normalization scale used for the retained masked AD channel $\widetilde{\mathbf{H}}^{\mathrm{AD,int}}$ computed using~\eqref{eq:architecture_rms}. We define the delay-cut energy ratio as $r_{\mathrm m}=\frac{ \|{\widetilde{\mathbf{H}}}^{\mathrm{AD,int}}\|_{\mathrm F}^2}
 {\|{\mathbf{H}}^{\mathrm{AD,int}}\|_{\mathrm F}^2}$, where ${\mathbf{H}}^{\mathrm{AD,int}}$ is the original masked AD domain input channel. The scale used for the clean channel is $ \rho_{\mathrm{out}}=\frac{\rho_{\mathrm{in}}}
 {\sqrt{p_{\mathrm{obs}}r_{\mathrm m}}}$, where $p_{\mathrm{obs}}$ is the observation fraction of the masked channel in SF domain defined in~\eqref{eq:observation_fraction}. 
This output scale approximates twice the clean retained-channel RMS under two assumptions: observed positions contain approximately a fraction $p_{\mathrm{obs}}$ of the SF channel energy, and delay truncation retains approximately all clean-channel energy.

For both channel feedback and interpolation, the head produces a normalized AD reconstruction. The BS restores its amplitude using the fed-back $\rho_{\mathrm{out}}$ before delay zero-padding and conversion to the SF domain, as described in Section~\ref{subsec:deployment}.

\textbf{LoS/NLoS identification:} To complement global channel features with amplitude information removed by RMS normalization, the CLS embedding is concatenated with the amplitude vector $\mathbf{a}$ derived from the unnormalized retained AD observation. Its five entries are the base-10 logarithms of the observation's maximum magnitude, RMS magnitude, mean magnitude, magnitude standard deviation, and 90th-percentile magnitude, which accompany CLS in the feedback. A two-hidden-layer multilayer perceptron (MLP) is used for classification. Training uses softmax cross-entropy, equivalent to the binary probability objective in~\eqref{eq:los_loss}. The same five-statistic amplitude input convention is used for localization and beam selection.

\textbf{User localization:} To learn the relation between global propagation features and UE position, an MLP network regresses two coordinates. Coordinate standardization places both output dimensions on comparable training scales. Training minimizes the coordinate-averaged squared error, corresponding to~\eqref{eq:localization_loss}, and the prediction is transformed back using training-set coordinate statistics. The reported positioning error remains the physical Euclidean distance in~\eqref{eq:position_error}.

\textbf{Single- and multi-user beam selection:} For a single UE, the MLP outputs $Q$ beam logits and is trained with cross-entropy against the reference label in~\eqref{eq:su_beam_target}. For $K$ UEs, beam choices must account for mutual interference. A shared user encoder first processes the CLS embedding of each UE, and a fusion MLP jointly processes the concatenated user features and amplitude statistics to account for group channel conditions. The head outputs $KQ$ logits, arranged as one $Q$-beam distribution per UE.

Cross-entropy encourages agreement with the reference beam labels but does not distinguish incorrect assignments by their achieved rate. We therefore combine it with a differentiable rate objective in the multi-user scenario, allowing training to reflect the performance consequence of a beam choice. Hard Gumbel-softmax sampling selects a codebook beam in the forward pass and supplies a continuous gradient approximation in the backward pass. The loss function for multi-user beam selection is
\begin{equation}
\mathcal{L}_{\mathrm{MU}}=\mathcal{L}_{\mathrm{CE}}
-\xi(e)\frac{N_c}{|\mathcal B|}
\sum_{\mathcal U\in\mathcal B}
R_{\mathcal U}^{\mathrm{MU}}(\widetilde{\mathbf q}_{\mathcal U}),
\label{eq:mu_training_loss}
\end{equation}
where $\mathcal{L}_{\mathrm{CE}}$ denotes the cross-entropy loss, $\widetilde{\mathbf{q}}_{\mathcal U}$ denotes the output beam indices, $e$ is training progress, and $\xi(e)$ gradually increases the rate-term weight.

\begin{table}[t]
\centering
\small
\setlength{\tabcolsep}{3pt}
\caption{Downstream head interfaces in the main framework.}
\label{tab:task_head_configuration}
\begin{tabular}{lll}
\toprule
Task & Head input & Output \\
\midrule
Channel feedback & $P\times D$ tokens & $2N_tN_\tau$ values \\
Interpolation & $P\times D$ tokens & $2N_tN_\tau$ values \\
LoS/NLoS & CLS + 5 statistics & 2 logits \\
Localization & CLS + 5 statistics & 2 coordinates \\
Single-user beams & CLS + 5 statistics & $Q$ logits \\
Multi-user beams & $K$ CLS + statistics & $KQ$ logits \\
\bottomrule
\end{tabular}
\end{table}

\section{Simulation Results}
\label{sec:results}
In this section, the simulation setup and results are introduced to demonstrate the performance of our proposed framework.

\subsection{Experimental Setup}
\label{subsec:experimental_setup}
The DeepMIMO dataset~\cite{lwma_deepmimo} is used to generate the pre-training and downstream channel data. LWM and LWMa are pre-trained on fifteen scenarios including about 800K channel samples: O1, Boston5G, ASU Campus, New York, Los Angeles, Chicago, Houston, Phoenix, Philadelphia, Miami, Dallas, San Francisco, Austin, Columbus, and Seattle. Downstream adaptation uses six separate scenarios (Denver, Indianapolis, Oklahoma, Fort Worth, Santa Clara, and San Diego) to evaluate the generalization of pre-trained representations to previously unseen environments. Compression networks and task heads are adapted separately in each downstream scenario. Table~\ref{tab:simulation_settings} summarizes the dataset parameters.

The proposed LWMa is compared with an LWM baseline in the SF domain using RMS normalization, denoted by LWM-RMS. Both employ the Perceiver IO-inspired embedding compression and recovery network. To validate the effectiveness of foundation models, single-task baselines combining CRNet-based channel feedback with task-specific downstream heads are compared. For identification, localization, and beam selection, the feedback network and task head are trained independently and cascaded at inference. Unless otherwise specified, plotted results are equally averaged over the six downstream scenarios. NMSE is averaged in the linear domain before conversion to decibels, and beam-selection rates are normalized within each scenario before averaging. Multi-user beam selection uses 2,500 training and 2,500 test user groups per scenario. Channel feedback uses 20~dB observations, and interpolation samples every fourth antenna and subcarrier. For simplicity, multi-user beam selection serves two users with identity digital precoding $\mathbf{F}_{\mathrm{BB},n}=\mathbf{I}$.

\begin{table}[t]
\centering
\caption{Dataset and Channel Settings}
\label{tab:simulation_settings}
\small
\begin{tabular}{@{}p{0.6\columnwidth}p{0.35\columnwidth}@{}}
\toprule
\textbf{Parameter} & \textbf{Setting} \\
\midrule
Downlink carrier frequency & $3.5$ GHz \\
Subcarrier spacing & $30$ kHz \\
Number of BS / UE antennas & $32/1$ \\
Number of subcarriers & $128$ \\
Retained delay bins for LWMa & $32$ \\
Feedback observation SNR & $20$ dB \\
Interpolation sampling strides & $4\times4$ \\
Beam codebook size & $64$ \\
Number of users for multi-user beam selection & $2$ \\
\bottomrule
\end{tabular}
\end{table}

For each LWMa model, a scenario-specific compressor is shared across tasks, while the foundation encoder remains fixed. Task heads are trained on uncompressed representations and then held fixed across compression budgets. Table~\ref{tab:compression_settings} lists the compression settings. The LoS/NLoS identification, localization, and beam-selection tasks additionally feed back the five entries of $\mathbf{a}$, while channel feedback and interpolation transmit the scalar $\rho_{\mathrm{out}}$. The feedback dimensions in Figs.~\ref{fig:embedding_recovery}--\ref{fig:recovered_tasks} count representation entries and exclude auxiliary amplitude information.

\begin{table}[t]
\centering
\caption{Compression and Adaptation Settings}
\label{tab:compression_settings}
\small
\begin{tabular}{ll}
\toprule
Parameter & Setting \\
\midrule
Embedding width $D$ & $64$ \\
Code dimension $Sr$ & $80,150,250,360$ \\
Slot configuration $(S,r)$ & $(20,4),(30,5),(50,5),(60,6)$ \\
Embedding training / fine-tuning & $200/200$ epochs \\
Fine-tuning weights $(\mu,\lambda)$ & $(0.25,0.2)$ \\
Task-head training & $120$ epochs \\
\bottomrule
\end{tabular}
\end{table}

\subsection{CLS-Driven Downstream Tasks}
\label{subsec:downstream_results}
Table~\ref{tab:cls_performance} compares the CLS-based foundation-model interfaces with the single-task baseline at a feedback dimension of 64. At the same feedback dimension, LWMa outperforms both LWM-RMS and the single-task baseline on all four averaged metrics. Its positioning error is reduced from 43.51 to 28.55~m relative to the single-task baseline, while the normalized single- and multi-user spectral efficiencies increase from 0.556 to 0.695 and from 0.389 to 0.613, respectively. These results demonstrate the utility of a shared global representation for multiple tasks without embedding compression.

\begin{table}[t]
\centering
\caption{Task Performance at a Feedback Dimension of 64}
\label{tab:cls_performance}
\begingroup
\footnotesize
\setlength{\tabcolsep}{3pt}
\renewcommand{\arraystretch}{1.08}
\begin{tabular*}{\columnwidth}{@{\extracolsep{\fill}}lrrr@{}}
\toprule
Task / Metric & LWMa & LWM-RMS & Single-task \\
\midrule
LoS/NLoS accuracy (\%) $\uparrow$ & 97.43 & 97.27 & 90.87 \\
Localization MPE (m) $\downarrow$ & 28.55 & 31.09 & 43.51 \\
Single-user normalized SE $\uparrow$ & 0.695 & 0.526 & 0.556 \\
Multi-user normalized SE $\uparrow$ & 0.613 & 0.428 & 0.389 \\
\bottomrule
\end{tabular*}
\par\vspace{2pt}
\begin{minipage}{\columnwidth}
\scriptsize
\textit{Note:} The results are mean values over six scenarios.
\end{minipage}
\endgroup
\end{table}

\subsection{Embedding Compression and Recovery}
\label{subsec:embedding_results}
\begin{figure}[t]
\centering
\includegraphics[width=\columnwidth]{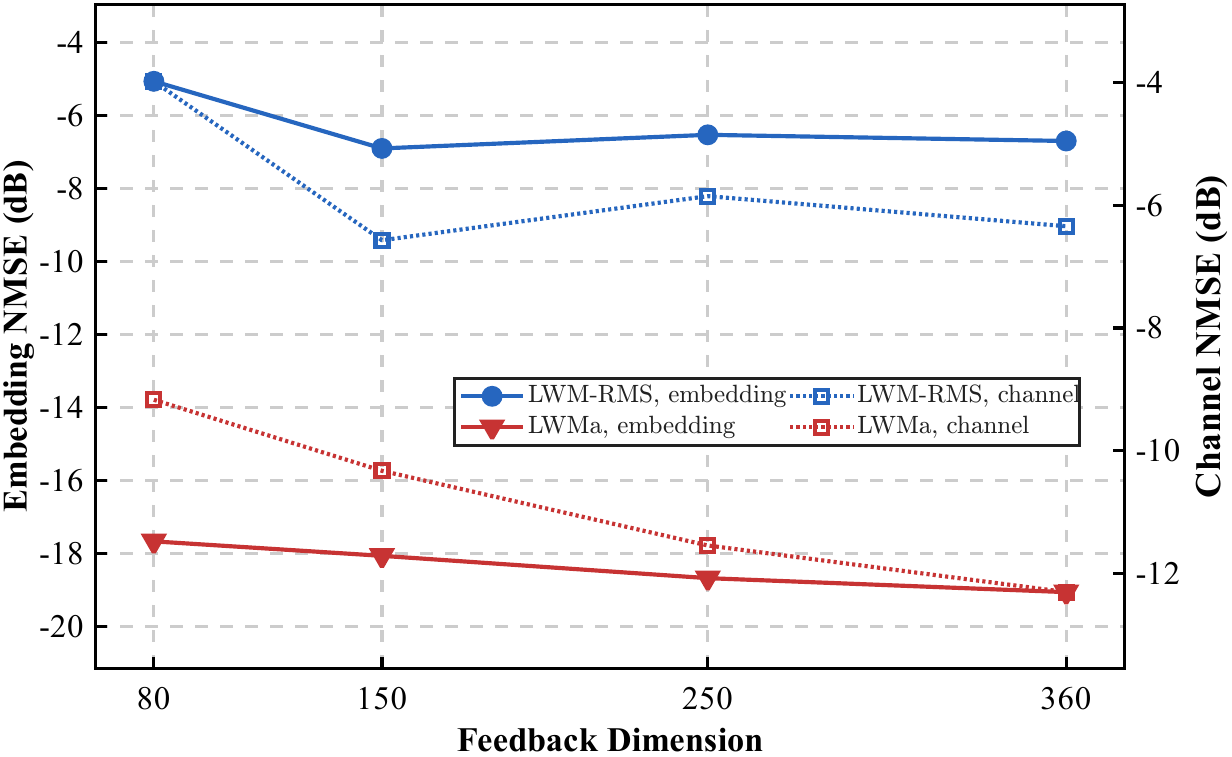}
\caption{Embedding recovery and frozen-decoder channel reconstruction averaged over six scenarios.}
\label{fig:embedding_recovery}
\end{figure}

First, we evaluate the proposed Perceiver IO-inspired embedding compression and recovery network with channel-quality-guided fine-tuning. Fig.~\ref{fig:embedding_recovery} shows that LWMa achieves average embedding NMSEs from $-17.67$ to $-19.06$~dB across the four budgets, compared with $-5.07$ to $-6.90$~dB for LWM-RMS. Increasing the budget from 80 to 360 also improves the frozen-decoder channel NMSE of LWMa. The recovered embeddings thus retain information useful for channel reconstruction. Note that the decoder metrics are measured in the truncated AD domain for LWMa and the full SF domain for LWM-RMS.

\begin{figure}[t]
\centering
\includegraphics[width=\columnwidth]{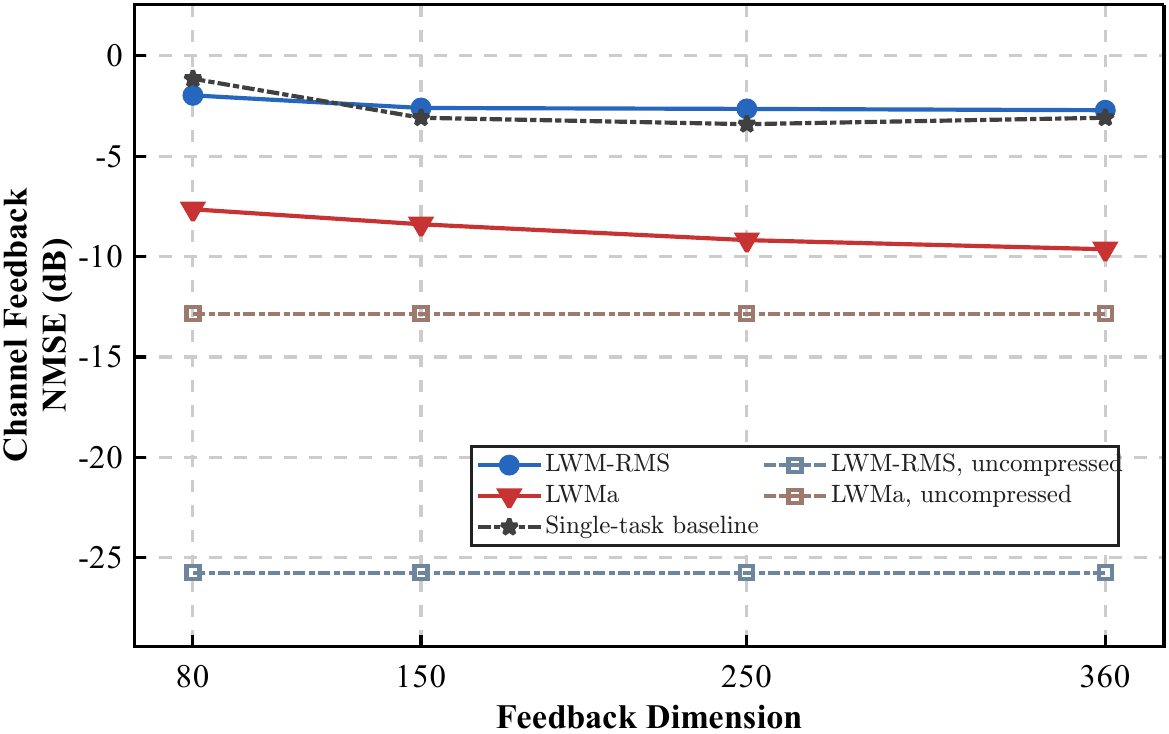}
\caption{Channel feedback performance averaged over six scenarios with 20~dB observation SNR.}
\label{fig:channel_feedback_results}
\end{figure}

\begin{figure}[t]
\centering
\includegraphics[width=\columnwidth]{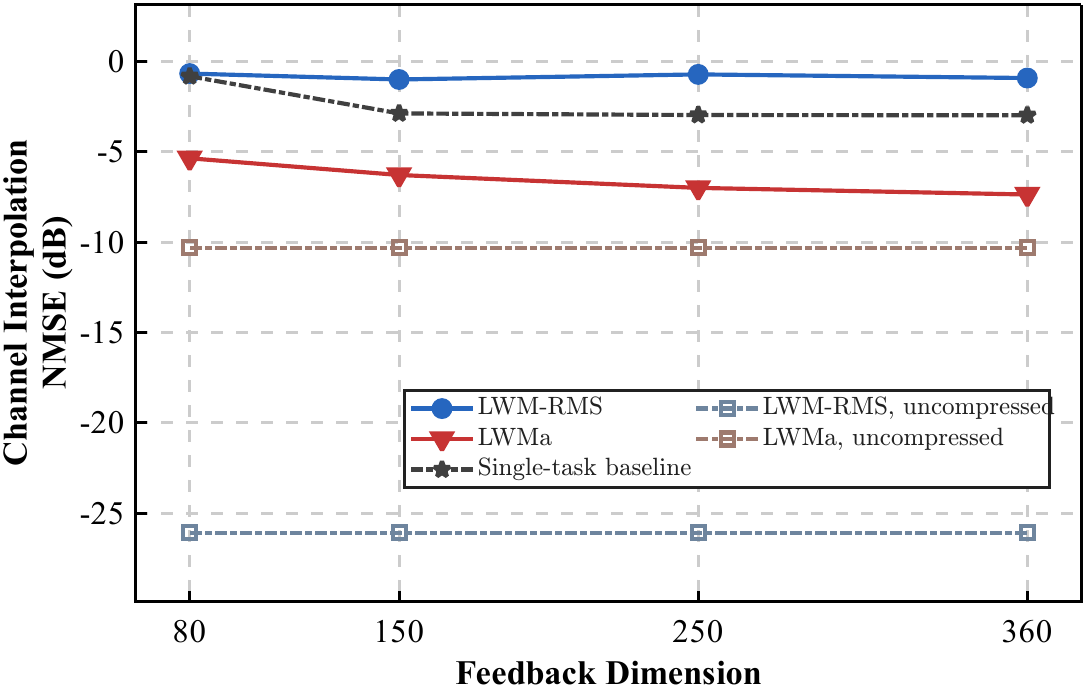}
\caption{Channel interpolation performance averaged over six scenarios with $4\times4$ sampling strides.}
\label{fig:channel_interpolation_results}
\end{figure}

Furthermore, we evaluate the channel feedback and interpolation driven by the embedding compression and recovery network, as shown in Figs.~\ref{fig:channel_feedback_results} and~\ref{fig:channel_interpolation_results}. At a budget of 360, LWMa achieves average feedback and interpolation NMSEs of $-9.64$ and $-7.37$~dB, improving on the single-task baseline by 6.55 and 4.38~dB, respectively. The same scenario-adapted compressor supports both tasks without retraining their heads, allowing the feedback budget to change with channel conditions. Table~\ref{tab:allscene_performance} further shows that LWMa outperforms both compressed LWM-RMS and the single-task baseline for both tasks at every tested budget in all six scenarios.

\begin{table*}[t]
\centering
\caption{Channel Recovery Performance Across Six Scenarios}
\label{tab:allscene_performance}
\begingroup
\footnotesize
\setlength{\tabcolsep}{3pt}
\renewcommand{\arraystretch}{1.08}
\begin{tabular*}{\textwidth}{@{\extracolsep{\fill}}ll*{12}{r}@{}}
\toprule
\multirow{2}{*}{Task / Metric} & \multirow{2}{*}{City} & \multicolumn{4}{c}{LWMa} & \multicolumn{4}{c}{LWM-RMS} & \multicolumn{4}{c}{Single-task} \\
 & & 80 & 150 & 250 & 360 & 80 & 150 & 250 & 360 & 80 & 150 & 250 & 360 \\
\midrule
\multirow{6}{*}{\shortstack[l]{Channel feedback\\NMSE (dB) $\downarrow$}} & San Diego & -8.75 & -9.61 & -10.00 & -9.91 & -4.40 & -4.73 & -6.13 & -5.18 & -1.29 & -3.04 & -3.61 & -3.24 \\
 & Santa Clara & -8.78 & -10.14 & -10.50 & -10.57 & -3.37 & -3.12 & -5.10 & -5.28 & -0.97 & -3.02 & -3.76 & -2.93 \\
 & Fort Worth & -8.61 & -8.93 & -9.36 & -9.37 & -1.11 & -1.19 & -1.72 & -1.13 & -0.97 & -3.45 & -3.45 & -3.17 \\
 & Indianapolis & -9.28 & -9.53 & -10.69 & -10.86 & -2.50 & -4.36 & -2.75 & -2.37 & -2.20 & -4.26 & -4.61 & -4.68 \\
 & Denver & -7.26 & -7.59 & -7.68 & -8.79 & -1.59 & -2.51 & -2.06 & -1.94 & -0.54 & -2.06 & -1.84 & -1.73 \\
 & Oklahoma & -4.92 & -6.04 & -7.86 & -8.77 & -0.20 & -1.06 & -0.55 & -2.06 & -1.09 & -3.00 & -3.69 & -3.29 \\
\addlinespace[3pt]
\multirow{6}{*}{\shortstack[l]{Channel interpolation\\NMSE (dB) $\downarrow$}} & San Diego & -5.89 & -7.15 & -7.37 & -7.40 & -2.00 & -2.00 & -1.47 & -2.03 & -0.85 & -2.79 & -3.01 & -3.06 \\
 & Santa Clara & -6.12 & -7.79 & -8.11 & -7.75 & -0.29 & -0.74 & -0.87 & -1.53 & -0.60 & -2.89 & -3.20 & -2.82 \\
 & Fort Worth & -6.03 & -7.05 & -7.60 & -7.55 & -0.41 & -0.57 & -1.36 & -0.26 & -0.61 & -3.04 & -2.93 & -3.25 \\
 & Indianapolis & -6.80 & -6.44 & -7.37 & -8.11 & -1.12 & -1.90 & -0.58 & -0.06 & -1.94 & -4.07 & -4.27 & -4.38 \\
 & Denver & -5.02 & -5.77 & -5.83 & -6.87 & -0.55 & -0.94 & -0.25 & -0.59 & -0.34 & -1.89 & -1.56 & -1.53 \\
 & Oklahoma & -3.27 & -4.40 & -6.17 & -6.69 & 0.02 & -0.17 & -0.01 & -1.42 & -0.79 & -2.87 & -3.31 & -3.37 \\
\bottomrule
\end{tabular*}
\endgroup
\end{table*}


\begin{figure*}[t]
\centering
\includegraphics[width=0.9\textwidth]{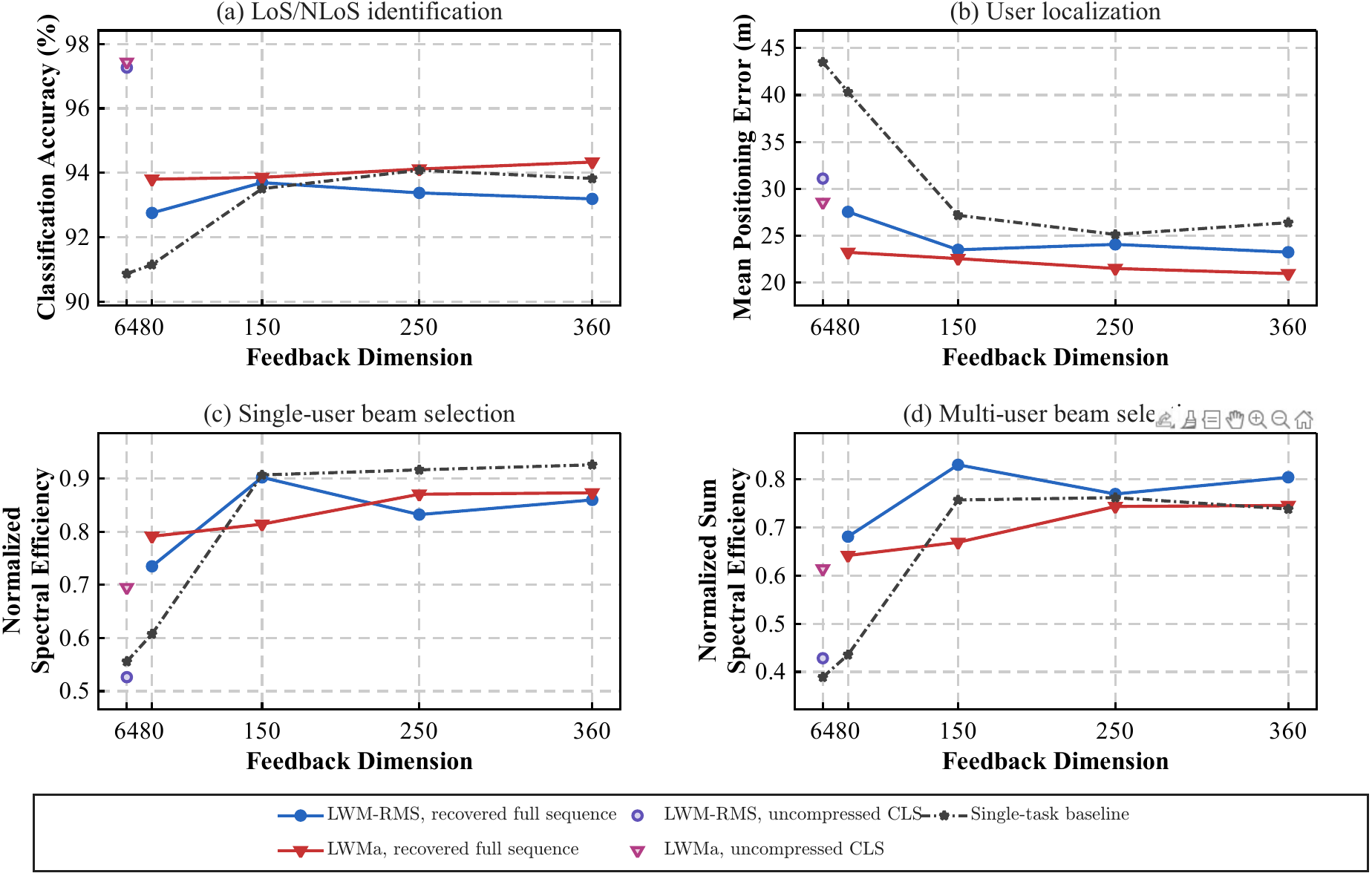}
\caption{Task performance using recovered embeddings, averaged over six scenarios. CLS references use 64-dimensional representations.}
\label{fig:recovered_tasks}
\end{figure*}

\label{subsec:compressed_transfer}
To evaluate how embedding compression generalizes across tasks, recovered full sequences are supplied to separately trained, frozen heads for identification, localization, and beam selection. The same scenario-specific compressors are reused without task-specific refinement. Fig.~\ref{fig:recovered_tasks} shows that LWMa improves average localization and both beam-selection metrics over its CLS representation at every tested budget. At 360 dimensions, the positioning error decreases from 28.55 to 20.96~m, while normalized single- and multi-user spectral efficiencies increase from 0.695 to 0.873 and from 0.613 to 0.745. These results demonstrate that the shared compressor preserves information useful beyond channel reconstruction.

\subsection{Deployment Complexity}
\label{subsec:deployment_complexity}
Table~\ref{tab:deployment_complexity} reports parameter counts and floating-point operations (FLOPs) for the model configurations. The UE hosts the foundation encoder and compression network, while the BS hosts the recovery network and task head. By exploiting the sparse angle-delay structure to truncate the delay dimension, LWMa shortens the embedding sequence and substantially reduces the computation required for feature extraction, compression, and recovery compared with LWM-RMS.
\begin{table}[t]
\centering
\caption{Deployment Complexity for All Six Tasks}
\label{tab:deployment_complexity}
\begingroup
\footnotesize
\setlength{\tabcolsep}{1.5pt}
\renewcommand{\arraystretch}{1.08}
\begin{tabular*}{\columnwidth}{@{\extracolsep{\fill}}llcrrrr@{}}
\toprule
\multirow{2}{*}{Method} & \multirow{2}{*}{Task} & \multirow{2}{*}{\shortstack{Feedback/\\Dimension}} & \multicolumn{2}{c}{UE} & \multicolumn{2}{c}{BS} \\
 & & & Param & FLOPs & Param & FLOPs \\
\midrule
\multirow{10}{*}{LWMa} & CF & compress/360 & 0.77 & 219.54 & 0.33 & 148.66 \\
 & CI & compress/360 & 0.77 & 219.54 & 0.33 & 148.66 \\
 & \multirow{2}{*}{ID} & CLS/64 & 0.59 & 197.54 & 0.30 & 0.60 \\
 &  & compress/360 & 0.77 & 219.54 & 4.69 & 50.16 \\
 & \multirow{2}{*}{LOC} & CLS/64 & 0.59 & 197.54 & 0.30 & 0.60 \\
 &  & compress/360 & 0.77 & 219.54 & 4.69 & 50.16 \\
 & \multirow{2}{*}{SU-BS} & CLS/64 & 0.59 & 197.54 & 0.33 & 0.66 \\
 &  & compress/360 & 0.77 & 219.54 & 4.72 & 50.22 \\
 & \multirow{2}{*}{MU-BS} & CLS/64 & 0.59 & 197.54 & 1.70 & 3.45 \\
 &  & compress/360 & 0.77 & 219.54 & 6.10 & 102.57 \\
\addlinespace[3pt]
\multirow{10}{*}{LWM-RMS} & CF & compress/360 & 0.90 & 1387.39 & 0.44 & 587.68 \\
 & CI & compress/360 & 0.90 & 1387.39 & 0.44 & 587.68 \\
 & \multirow{2}{*}{ID} & CLS/64 & 0.66 & 1352.78 & 0.30 & 0.60 \\
 &  & compress/360 & 0.90 & 1387.39 & 17.38 & 191.86 \\
 & \multirow{2}{*}{LOC} & CLS/64 & 0.66 & 1352.78 & 0.30 & 0.60 \\
 &  & compress/360 & 0.90 & 1387.39 & 17.38 & 191.86 \\
 & \multirow{2}{*}{SU-BS} & CLS/64 & 0.66 & 1352.78 & 0.33 & 0.66 \\
 &  & compress/360 & 0.90 & 1387.39 & 17.42 & 191.92 \\
 & \multirow{2}{*}{MU-BS} & CLS/64 & 0.66 & 1352.78 & 1.70 & 3.45 \\
 &  & compress/360 & 0.90 & 1387.39 & 18.85 & 385.98 \\
\addlinespace[3pt]
\multirow{6}{*}{Single-task} & CF & -/360 & 0.74 & 1.79 & 0.74 & 7.22 \\
 & CI & -/360 & 0.74 & 1.79 & 0.74 & 7.22 \\
 & ID & -/360 & 0.74 & 1.79 & 1.38 & 16.07 \\
 & LOC & -/360 & 0.74 & 1.79 & 1.55 & 12.93 \\
 & SU-BS & -/360 & 0.74 & 1.79 & 1.57 & 14.17 \\
 & MU-BS & -/360 & 0.74 & 1.79 & 2.95 & 49.34 \\
\bottomrule
\end{tabular*}
\par\vspace{2pt}
\begin{minipage}{\columnwidth}
\scriptsize
\textit{Note:} Param and FLOPs are reported in millions ($10^6$). CF: channel feedback; CI: channel interpolation; ID: LoS/NLoS identification; LOC: user localization; SU-BS/MU-BS: single-/multi-user beam selection.
\end{minipage}
\endgroup
\end{table}

\subsection{Split-Inference Evaluation}
\label{subsec:split Indianapoli}
To alleviate the computational burden at the UE, the deployment can also adopt split inference, which moves the final three encoder layers of LWMa to the BS. The UE compresses the output after layer nine, and the BS recovers this intermediate embedding before executing the remaining layers and task head. Identification, localization, and beam selection use the recovered full sequence in this comparison.

\begin{figure*}[t]
\centering
\includegraphics[width=\textwidth]{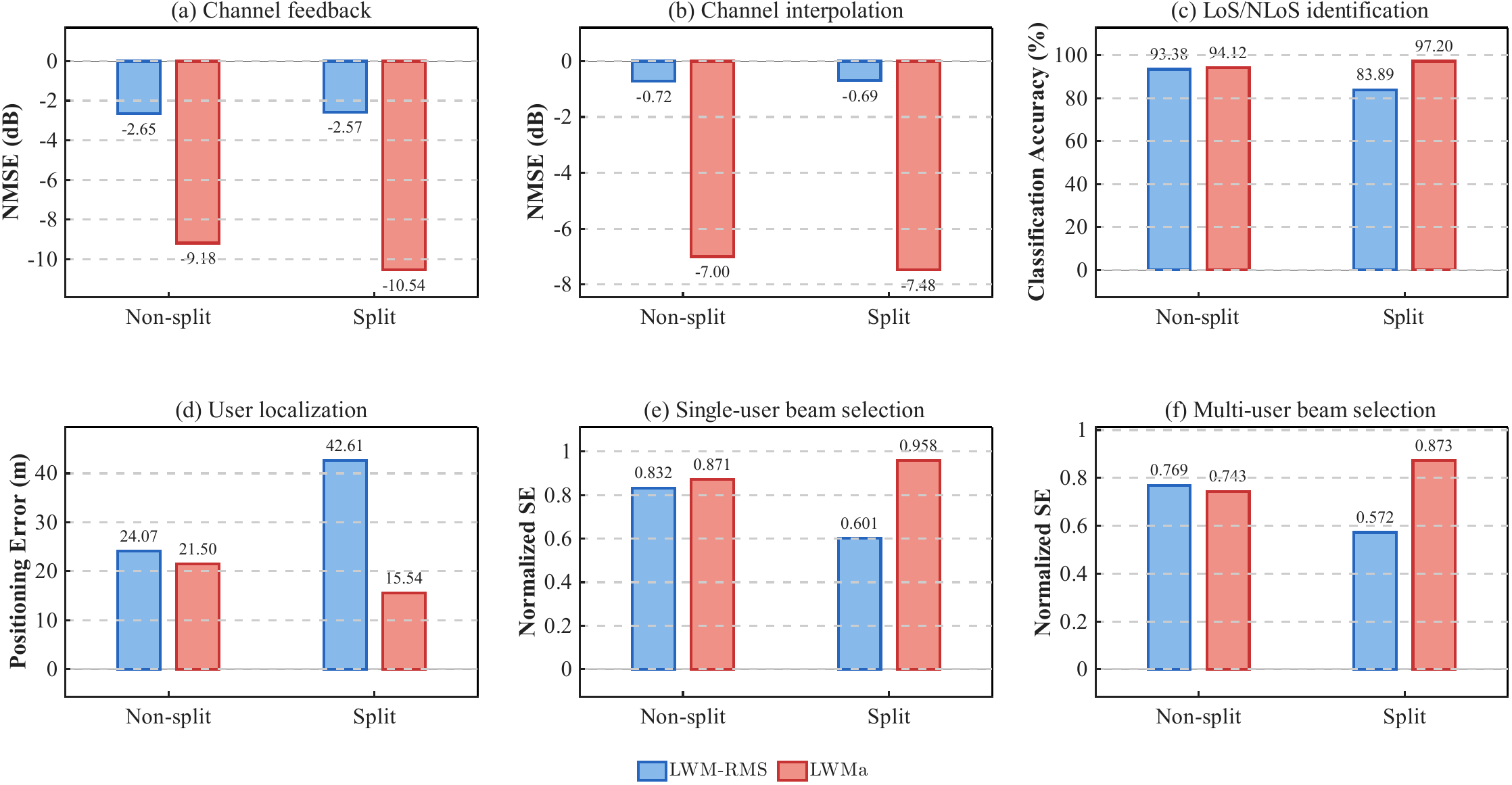}
\caption{Task performance with and without layer-nine split inference at a feedback dimension of 250, averaged over six scenarios. Identification, localization, and beam selection use the recovered full sequence.}
\label{fig:split_results}
\end{figure*}

Fig.~\ref{fig:split_results} compares the two deployments at a feedback dimension of 250 with the embedding compression and recovery network, averaged over six scenarios. LWMa improves all six averaged metrics in the split-inference deployment. Feedback and interpolation NMSE decrease by 1.36 and 0.48~dB, respectively, and the positioning error decreases by 27.7\%. Identification accuracy increases from 94.12\% to 97.20\%, while normalized single- and multi-user spectral efficiencies rise from 0.871 to 0.958 and from 0.743 to 0.873. LWM-RMS degrades under the same split configuration. These results support the use of LWMa intermediate representations to move encoder computation from the UE to the BS while maintaining useful task information.

\section{Conclusion}
\label{sec:conclusion}
This paper proposed an LWMa-based distributed framework that supports multiple tasks through compressed embedding feedback or direct global-embedding feedback. Angle-delay processing in the LWMa network reduces the representation length and computational cost, while scenario-specific compression and channel-quality-guided training preserve channel information in recovered embeddings for independently trained heads. Across downstream scenarios, LWMa improves channel feedback and interpolation accuracy over the evaluated baselines. Recovered sequences also support identification, localization, and beam selection, with gains over the global embedding depending on the task and scenario. The evaluated split deployment further demonstrates the potential to move encoder computation from the UE to the BS while retaining useful task performance.


\end{document}